\documentclass[11pt]{article}

\usepackage[title]{appendix}

\usepackage[margin={.8in}]{geometry}

\usepackage{rotating, lscape}
\usepackage{graphicx, color, subfig}
\usepackage{url}

\usepackage{threeparttable, booktabs, longtable, threeparttablex, tabularx}

\usepackage{natbib}
\usepackage{authblk}

\usepackage[sc]{mathpazo}
\usepackage{courier}

\usepackage[utf8]{inputenc}
\usepackage[T1]{fontenc}

\usepackage{setspace}
\usepackage{footmisc}
\usepackage{amssymb,amsmath,mathrsfs,amsthm, bbm}

\usepackage{microtype}

\usepackage{float}
\usepackage[hidelinks]{hyperref}
\usepackage{multirow}
\usepackage{pdflscape}
\usepackage[capitalize]{cleveref}

\newcommand{\E}{\mathbb{E}}
\newcommand{\Var}{\text{Var}}
\newcommand{\Cov}{\text{Cov}}

\newcommand{\pub}{\text{pub}}

\newcommand{\FDR}{\text{FDR}}

\definecolor{ChadBlue}{rgb}{.1,.1,.5}  

\begin{document}

\title{Publication Bias in Asset Pricing Research\footnotetext{E-mails:  andrew.y.chen@frb.gov and tom.zimmermann@uni-koeln.de. We thank Alec Erb and Arne Rodloff for excellent research assistance.  We thank Vadim Elenev, Andrei  Gonçalves, Amit Goyal, Dalida Kadyrzhanova, Ben Knox, Dino Palazzo, Lasse Pedersen, Michael Palumbo, Mihail Velikov, Fabian Winkler, and seminar participants at the Federal Reserve Board, Johns Hopkins University, KAIST / Korea University Virtual Seminar,  Renmin University of China, and Texas A\&M for helpful comments. These views are not necessarily those of the Federal Reserve Board or the Federal Reserve System. Replication code is found at \url{https://github.com/chenandrewy/PBR} }}

\author[1]{Andrew Y. Chen}
\author[2]{Tom Zimmermann}

\affil[1]{Federal Reserve Board}
\affil[2]{University of Cologne}

\maketitle

\begin{abstract}

\noindent \normalsize Researchers are more likely to share notable findings. As a result, published findings tend to overstate the magnitude of real-world phenomena. This bias is a natural concern for asset pricing research, which has found hundreds of return predictors and little consensus on their origins.  

Empirical evidence on publication bias comes from large scale meta-studies.  Meta-studies of cross-sectional return predictability have settled on four stylized facts that demonstrate publication bias is \emph{not} a dominant factor: (1) almost all  findings can be replicated, (2) predictability persists out-of-sample, (3) empirical $t$-statistics are much larger than 2.0, and (4) predictors are weakly correlated.  Each of these facts has been demonstrated in at least three meta-studies.      

Empirical Bayes statistics turn these facts into publication bias corrections.  Estimates from three meta-studies find that the average correction (shrinkage) accounts for only  10 to 15 percent of  in-sample mean returns and that the risk of inference going in the wrong direction (the false discovery rate) is less than 10\%. 

Meta-studies also find that $t$-statistic hurdles exceed 3.0 in multiple testing algorithms and that returns are 30 to 50 percent weaker in alternative portfolio tests.  These facts are easily misinterpreted as evidence of  publication bias effects. We clarify these misinterpretations and others, including the conflating of  ``mostly false findings'' with ``many insignificant findings,'' ``data snooping'' with ``liquidity effects,'' and ``failed replications'' with ``insignificant ad-hoc trading strategies.''

Meta-studies outside of the cross-sectional literature are rare. The four facts from cross-sectional meta-studies provide a framework for future research.  We illustrate with a preliminary re-examination of  equity premium predictability.

\end{abstract}

\medskip


\newpage

\doublespacing

\section{Introduction}\label{sec:intro}
What's the point of publishing something if it's not worth noting?   In most forms of publishing, authors are motivated to write notable content.  This motive means that seemingly-notable content is over-represented in any set of published writings, and thus any set of published writings exaggerates the truth.  Scientific publishing is no exception.

Science is unique in its precise definition of notable content.  Even before \citet{fisher1925statistical}, researchers have said a finding is notable if it is at least ``5\% significant''  (\citet{stigler2008fisher}).   We refer to the bias toward statistically significant results, as well as the effects of this bias, as ``publication bias.''

Publication bias may seem like a very slippery issue.  But its essence is found in a simple decomposition.  Reported effects (e.g. regression coefficients) are the sum of the true effect and errors:
\begin{align}\label{eq:intro}
    \text{Reported effect} = \text{True effect} + \text{Author error} + \text{Sampling error}
\end{align}
where sampling errors are statistical noise and author errors represent non-statistical problems unrelated to economics (intentional and unintentional mistakes).  In an ideal world, both errors are on average zero, so the reported effect would be equal to the true effect in expectation.  But under publication bias, only large reported effects are published.  This bias selects for large author errors and large sampling errors, implying 
\begin{align}\label{eq:intro2}
    \E(\text{Reported effect} |\text{Published}) >
    \E(\text{True effect} | \text{Published}),
\end{align}
that is, published reported effects are upward biased relative to the true effect.  The gap between the LHS and RHS of Equation \eqref{eq:intro2} measures the magnitude of publication bias.

To correct for publication bias, we just need to estimate
\begin{align}\label{eq:intro3}
  \E(\text{Author error} |\text{Published}) +
    \E(\text{Sampling error} | \text{Published}).    
\end{align}
Subtracting these error terms from $ \E(\text{Reported effect} |\text{Published})$ leads to an unbiased estimate of the true effect conditional on publication.  Meta-studies with large collections of published findings can provide estimates of these error terms.

Terms like  ``the multiple testing problem,''  ``data-snooping,'' ``p-hacking,'' ``the file drawer problem,'' and ``researcher degrees of freedom''  fill out the details.  In the end, these problems boil down to selection bias and the need to correct for Equation \eqref{eq:intro3}.

This article reviews research on publication bias in asset pricing.  Empirical evidence is abundant in cross-sectional return predictability, so we focus our review in this sub-field.  We begin with four facts that emerge from these studies (Section \ref{sec:facts}).  We then describe shrinkage corrections and false discovery rate  estimates, which pin down publication bias at 10\% to 15\% of in-sample returns (Section  \ref{sec:gap}).   
We square these results with other facts  like ``anomaly decay''  and clarify related misinterpretations regarding the prevalance of ``false findings''  in  Section \ref{sec:mis}. We end by illustrating how our facts provide a framework for understanding publication bias more broadly, by examining evidence from equity premium prediction (Section \ref{sec:conclusion}).

Before we continue, we note that there are other forms of publication bias that we do not cover.  Our notion of publication bias follows the common definition of selection for notable results (\citet{delong1992all}; \citet{brodeur2016star}; \citet{andrews2019identification}; etc).  But researchers may also be biased toward writing papers that confirm the existing paradigm (\citet{kuhn1962structure}; \citet{akerlof2018persistence}).   To our knowledge, the only empirical evidence on this kind of publication bias in finance is \citet{rubin2021systematic}, who uses a natural experiment to show that a significant share of paper citations in top tier finance journals exist to curry favor with influential scholars.   

\section{Four Facts about Publication Bias in the Cross-Section of Returns}\label{sec:facts}

Cross-sectional predictability refers to the idea that some stocks have higher average future returns than others.  These differences in expected returns are typically  documented by sorting stocks into portfolios based on some firm-level characteristic, and then examining the performance of going long-short the extreme portfolios.  A series of meta-studies re-examine dozens, or even hundreds of published cross-sectional predictors (e.g. \citet{green2013supraview};      \citet{Mclean2016Does}; \citet{ChenZimmermann2021}).  

This literature finds four salient facts:
\begin{enumerate}
\item Almost all predictability findings can be replicated
\item Predictability persists out-of-sample
\item Empirical t-stats are much, much larger than 2.0
\item Predictors are weakly correlated
\end{enumerate}
Each of these facts has been documented by at least three different meta-studies.

The importance of these facts can be seen in Equation \eqref{eq:intro}. High replicability (Fact \#1) rules out author error as a meaningful influence.  Out-of-sample persistence (Fact \#2) limits the impact of sampling error, as sampling error should vanish out-of-sample.  Very large t-stats (Fact \#3) also limit the impact of sampling error, as they imply sampling error is small.  Weak correlations (Fact \#4) show that the zoo of predictors is not redundant, and also validates assumptions used in the multiple testing statistics.

We illustrate these facts using the \citet{ChenZimmermann2021} (CZ22) dataset.  This dataset consists of replications of hundreds of variables that were claimed to be related to cross-sectional predictability.  It  includes the vast majority of predictors found in previous meta-studies (\citet{Mclean2016Does}; \citet{green2017characteristics}; \citet{Hou2020Replicating}).  Unlike other meta-studies, we compare our replications to hand-collected findings from the original papers.  We also post data, code, and detailed documentation at \url{https://www.openassetpricing.com/}, and update these data annually, fixing errors and occasionally adding new predictors.  The version used in this review is the March 2022 release, which contains 207 cross-sectional predictors published in 140 papers. 

The dataset includes many portfolio implementations of these 207 predictors.  For this review, we focus on raw long-short portfolios constructed following the methods in the  original papers.  Code to replicate the results in this review are found at \url{https://github.com/chenandrewy/PBR}.

\subsection{Fact \#1: Almost all predictability findings can be replicated}\label{sec:facts-rep}

\begin{figure}[h!]
    \caption{
    {Replicated vs Original Paper t-stats.} Each marker represents a different cross-sectional predictor from the CZ22 dataset. t-stats test the hypothesis that the mean return on a long-short strategy that trades on the predictor in the original sample period is zero. We show only predictors for which the original papers provide raw long-short t-statistics.   Solid line is the 45 degree line.  On average, CZ22's replicated t-stats match the original t-stats.  These replications can surely be improved with more economist-hours.
}
    \label{fig:rep_vs_hand}
    \vspace{.10in}
    \centering     
                                \includegraphics[width=5in]{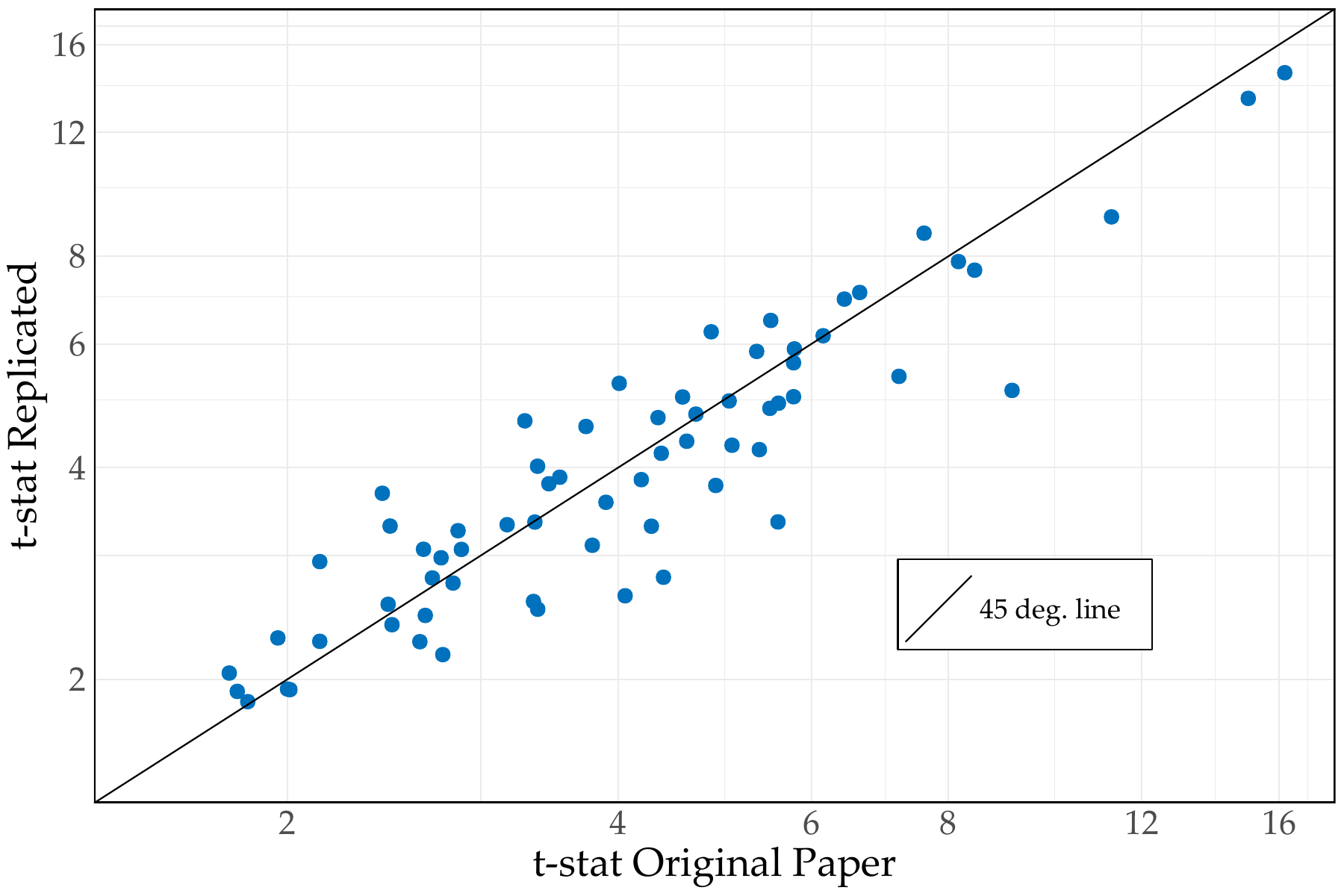}
                                                               
\end{figure}

\citet{Mclean2016Does} and \citet*{Jensen2022Is} find that most of their implementations of published cross-sectional predictors are statistically significant.  CZ22 go further and show that nearly \emph{all} findings can be replicated.\footnote{In economics, psychology, and biomedical research, there is little consensus on what ``replication'' means (\citet{hamermesh2007replication}; \citet{open2015estimating}; \citet*{patil2019visual}; \citet{welch2019reproducing}). We avoid jargon and use the dictionary definition:  ``the act of making or doing something again in exactly the same way'' (Cambridge Dictionary).}         Figure \ref{fig:rep_vs_hand} illustrates this result.

The figure plots CZ22's replicated t-stats  against t-stats hand-collected from the original papers.  The markers line up nicely along the 45 degree line, showing that cross-sectional predictability can not only be replicated qualitatively, but quantitatively.

The replicated t-stats deviate somewhat from the originals.  CZ22 aim to cover all predictors in  four previous meta-studies, leading to a list of 319 characteristics from 153 papers.  These 319 characteristics are replicated with a team of just two economists, so each replication can surely be improved with more economist-hours.   But considering this limitation, the match is remarkable.



 
This result bounds the effects of ``author error'' in Equation \eqref{eq:intro3}.   If author error was a big problem, then the markers would lie far below the 45 degree line. Instead, Figure \ref{fig:rep_vs_hand} shows that author error is on average close to zero.  In other words, the numbers in the original papers are a fairly unbiased estimate of what a researcher who attempts to replicate the original findings would obtain.

In contrast, \citet*{Hou2020Replicating} find that ``most anomalies fail to replicate.'' This is commonly reconciled with other meta-studies by noting that Hou et al.'s long-short strategies de-emphasize microcap stocks.   However, this reconciliation is not correct.  The proper reconciliation is misclassification of failed replications: only about 26\% of Hou et al.'s long-short strategies were shown to be clearly statistically significant in the original papers (\citet{ChenZimmermann2021}; see also \citet{Jensen2022Is}).  We discuss this and other misinterpretations in Section \ref{sec:mis}.


\subsection{Fact \#2: Predictability persists out-of-sample}\label{sec:facts-oos}

If predictability was entirely due to publication bias, then it should
vanish immediately after the original samples end. But \citet{Mclean2016Does} (MP) find nearly the opposite pattern.  

MP replicate 97 published predictors and compare their returns from the original sample periods to their returns between the end of the original sample and the publication date. Returns decline by only 26\%. In other words, the in-sample returns are still 74\% there.   A similar out-of-sample decline is documented in \citet{Jacobs2018And} and  \citet{chen2020publication}.

\begin{figure}[h!]
    \caption{
                                      Predictability persists out-of-sample. Long-short returns are scaled so that the mean in-sample return is 100 bps per month for ease of interpretation. Blue shows the mean return pooled across all in-sample observations.  Gray shows the pooled mean return using data from the first three years after the original samples end.  Distributions indicate sampling uncertainty via cluster bootstrap, which draws all returns in the same month simultaneously.  74\% of returns persist in these out-of-sample periods.
                                }
    \label{fig:MPbootstrap}
    \vspace{.10in}
    \centering     
                                \includegraphics[width=4in]{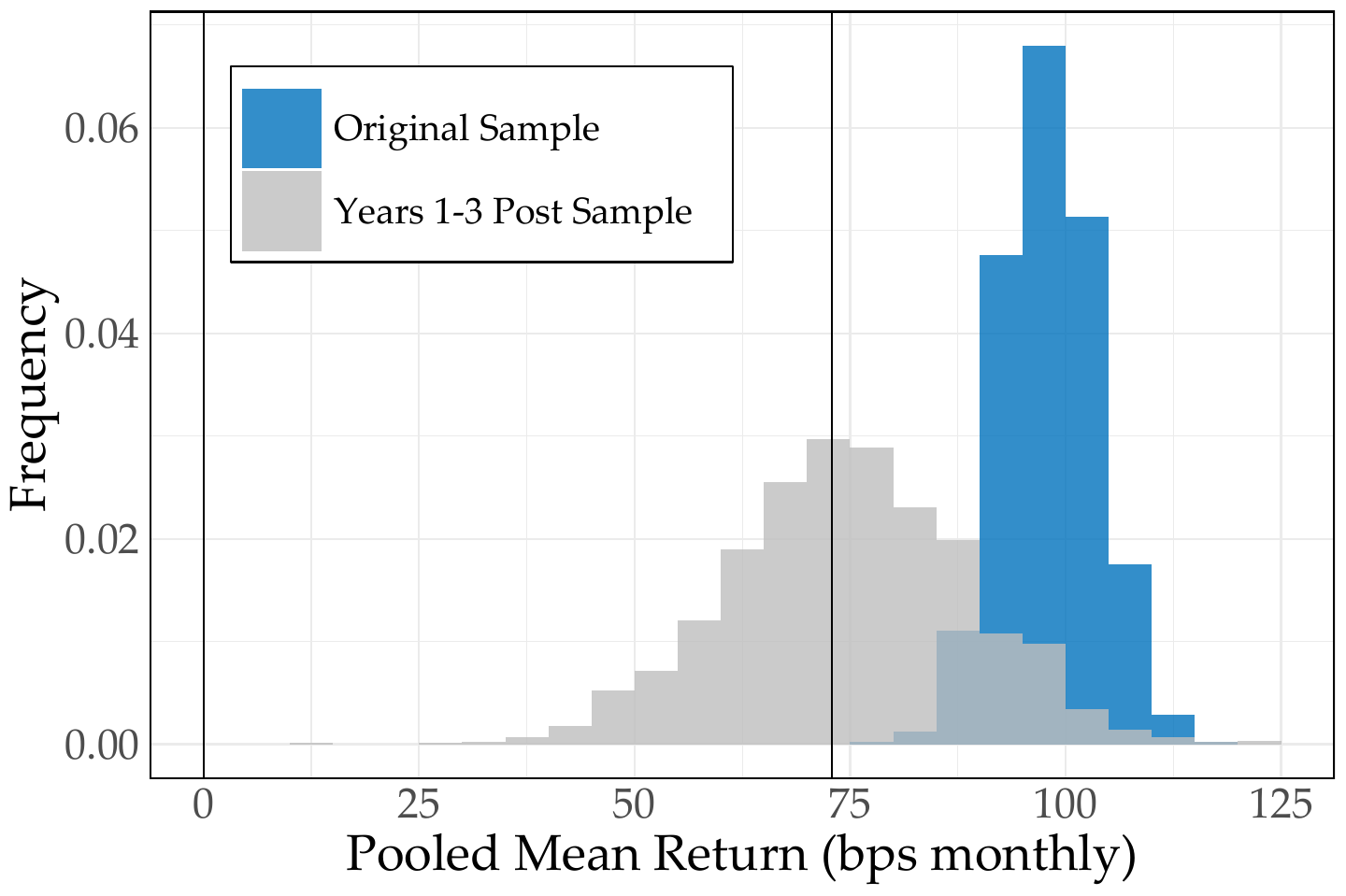}
                                                               
\end{figure}

Figure \ref{fig:MPbootstrap} replicates MP's result using the CZ22 dataset.  Long-short returns are scaled so that the mean in-sample return is 100 bps per month for ease of interpretation.  In the first 3 years after the end of the original samples, the mean return is still 74 bps per month (vertical line).  These out-of-sample periods are short, but large scale replications like MP's still show statistical significance.  This significance is seen in the bootstrapped distribution of the out-of-sample mean  (grey bars).  After accounting for sampling uncertainty, these out-of-sample returns are  still far from zero, and quite close to in-sample returns (blue bars).

MP's 26\% decline places an upper bound on the effects of  ``sampling error'' in Equation \eqref{eq:intro3}.    This small decline is a tough challenge for believers in the dominance of publication bias.  As argued by MP, this 26\% decline is an upper bound because investors should learn from the academic publications, and may trade away (or avoid creating) the documented return patterns (see also \citet{marquering2006disappearing}).   The actual effects of sampling error may be much smaller than 26\%.  

Returns far from the original sample periods have a larger decay of roughly 50\% (\citet{Mclean2016Does}; \citet{Linnainmaa2018The}; \citet{Jacobs2020Anomalies}; \citet{Jensen2022Is}).   Since publication bias should show up immediately after the samples end, the remaining decay is due to real changes in expected returns (see Equation \eqref{eq:intro}).  These changes are natural given that many, if not most of the original papers argue for mispricing-based explanations (\citet{rosenberg1985persuasive}; \citet{jegadeesh1993returns}; etc).  We return to this issue in Section \ref{sec:mis}.

\subsection{Fact \#3: Empirical t-stats are much larger than 2.0}\label{sec:facts-gap}
The common story for publication bias goes something like this: even if there's no predictability, you'll find 5\% significance if you just p-hack $1/0.05 = 20$ strategies.  Since 5\% significance corresponds to a t-stat of 2.0, this suggests it's quite possible that t-stats near 2.0 are due to publication bias.  

But the t-stats found in the literature are much larger than 2.0.   Table \ref{tab:t-too-big} provides the details, using the distribution of t-stats found in the CZ22 data.  The majority of t-stats exceed 3.0, and plenty of t-stats exceed 4.0.  There are even predictors with t-stats that exceed 8.0.  Similar results are found in hand-collected t-stats (\citet{green2013supraview}; \citet{chen2020publication}).

\begin{table}[!htbp]
      \caption{Published t-stats are much  larger than 2.0.}
       ``Published and Replicated'' are from CZ22's replications of 207 predictors. ``Systematically Data-Mined'' are from \citet{yan2017fundamental}'s 18,113 equal-weighed strategies mined by sorting stocks on simple functions of 240 accounting variables.  The bootstrap draws de-meaned returns from the CZ22 data, with returns in the same month drawn simultaneously to control for correlations.  Large t-stats are thousands of times more common than implied by the null of no predictability.
      \label{tab:t-too-big}  
\begin{center}

\begin{tabular}{lrrrrrrr}
\toprule
  & \multicolumn{7}{c}{t-stat minimum} \\
  & 2.0 & 3.0 & 4.0 & 5.0 & 6.0 & 7.0 & 8.0 \\
\cmidrule{2-8}  &   &   &   &   &   &   &  \\
  & \multicolumn{7}{c}{(a) Number of Predictors that Meet Minimum} \\
\cmidrule{2-8}Published and Replicated & 183 & 121 & 74 & 48 & 26 & 18 & 12 \\
Systematically Data-Mined &                         5,464  &          2,837  &          1,522  &             832  &             374  &             185  &               76  \\
  &   &   &   &   &   &   &  \\
  & \multicolumn{7}{c}{(b) Percent of Signals that Meet Minimum} \\
\cmidrule{2-8}Published and Replicated & 88.4058 & 58.4541 & 35.7488 & 23.1884 & 12.5604 & 8.6957 & 5.7971 \\
Systematically Data-Mined & 30.1662 & 15.6628 & 8.4028 & 4.5934 & 2.0648 & 1.0214 & 0.4196 \\
\midrule
  &   &   &   &   &   &   &  \\
  & \multicolumn{7}{c}{(c) Percent of Signals Implied by Null (Significance Level)} \\
\cmidrule{2-8}Standard Normal & 4.5500 & 0.2700 & 0.0063 & 0.0001 & 0.0000 & 0.0000 & 0.0000 \\
Bootstrap & 4.6115 & 0.3034 & 0.0107 & 0.0002 & 0.0000 & 0.0000 & 0.0000 \\
\bottomrule
\end{tabular}%

\end{center}
\end{table} 

While it would take only  about $1/0.05 = 20$ draws to generate a t-stat greater than 2.0, it would take  $1/0.000063 \approx 16,000$ draws to generate just one of the 74 t-stats that exceeds 4.0.  For the 26 t-stats that exceed 6.0, the significance is so strong that it can't be seen in the table.  It would take $1/0.000000002 = 50$ billion draws to generate just one of these t-stats.  It is physically impossible that these 26 t-stats are found in the literature if there is no predictability (\citet{chen2021limits}).  

These thought experiments illustrate the limits of ``sampling error'' in Equation \eqref{eq:intro}.  Sampling error can only explain t-stats that are  close to the null of no predictability.  t-stats far from the null, then, must be due to positive expected returns.  And while publication bias may distort the magnitude of published t-stats, this distortion would have to be very extreme to account for the gap seen in Table \ref{tab:t-too-big}.  Moreover, publication bias does not distort the empirical probabilities found in atheoretical data-mining experiments   (\citet{yan2017fundamental}; \citet{Chordia2020Anomalies}). As seen in Panel (b), even atheoretical data mining generates large t-stats thousands of times too frequently to be due to luck. 

Multiple testing statistics convert this intuition into publication bias corrections.  Shrinkage estimates imply that sampling error in Equation \eqref{eq:intro3} is only 10\% to 15\% of in-sample returns.  False discovery rate estimates  imply that the risk of expected returns being zero or negative is less than 10\%. These small publication bias effects are found in models estimated by three independent meta-studies (\citet{harvey2016and}; \citet{chen2020publication}; \citet{Jensen2022Is}).   We explain these statistics in detail in Section \ref{sec:gap}.    

In contrast, \citet{harvey2016and} famously ``argue that most findings in financial economics are likely false.''  Despite this language, their estimates imply a  false discovery rate of less than 10\% (see Section \ref{sec:gap-lit} below).  This confusion is also seen in Harvey et al's conclusion, which  mixes the terms ``mostly false findings'' with  ``many insignificant findings.''  It's easy to confuse significance with truth in standard hypothesis testing  (\citet{wasserstein2016asa}).  It's even easier to mix up the language in multiple testing statistics.  We discuss this issue further in Section \ref{sec:mis}.

\subsection{Fact \#4: Predictors are weakly correlated }\label{sec:facts-cor}
A common view is that the zoo of predictors are just hacked copies of a handful of real predictors like value and momentum.  But the data show  the predictors are quite diverse.

\begin{figure}[h!]
    \caption{Predictors are weakly correlated.  Data consists of monthly long-short returns from the CZ22 dataset.  All returns are signed to have positive in-sample means. Almost all correlations are less than 0.5 and 60 principal components are required to span 90\% of total variance, indicating a zoo of distinct predictors.}
    \label{fig:cor}
    \vspace{.15in}
    \centering
                               
            \subfloat[Distribution of Correlations]{\includegraphics[width=0.47\textwidth]{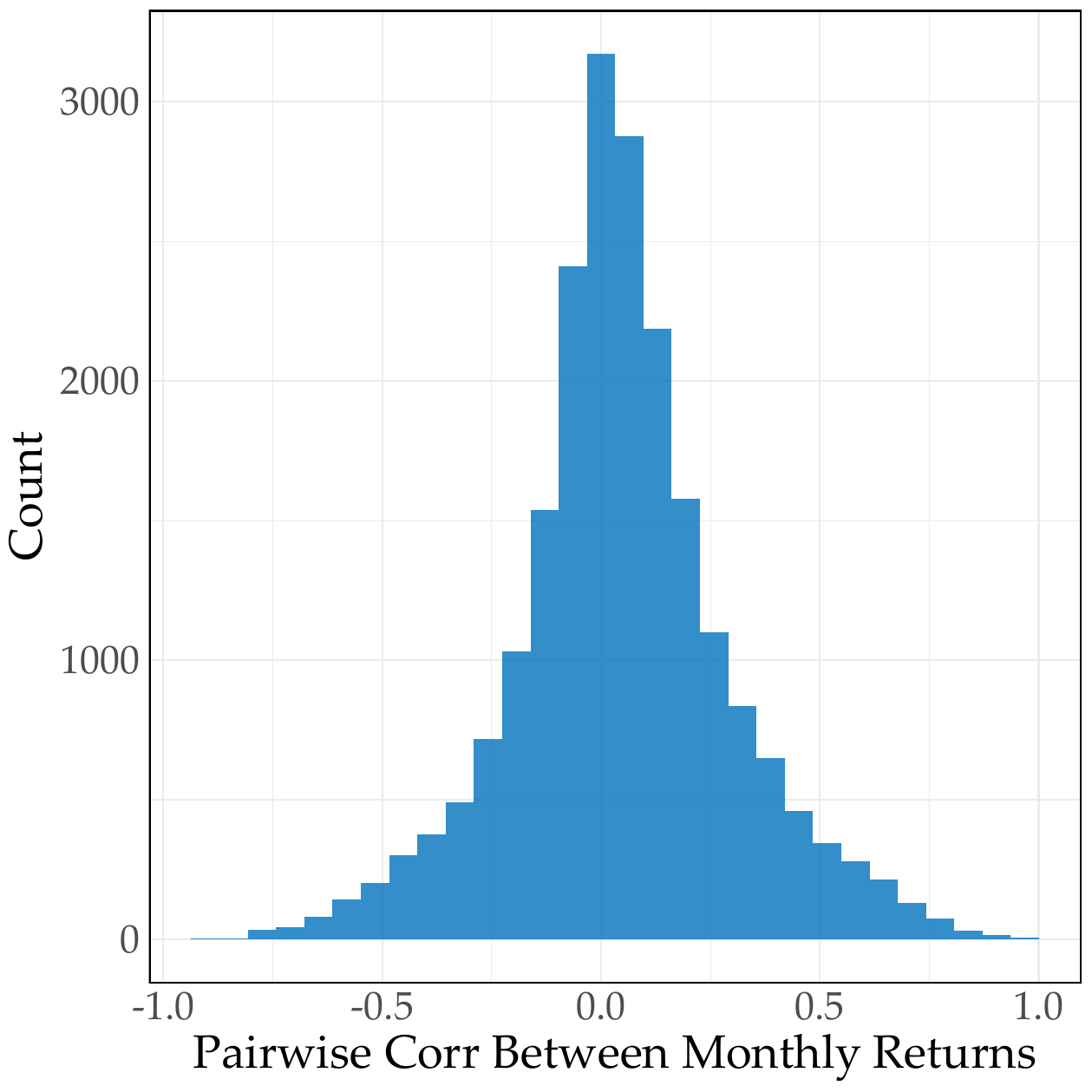}}
      \qquad
            \subfloat[Principal Component Analysis]{\includegraphics[width=0.47\textwidth]{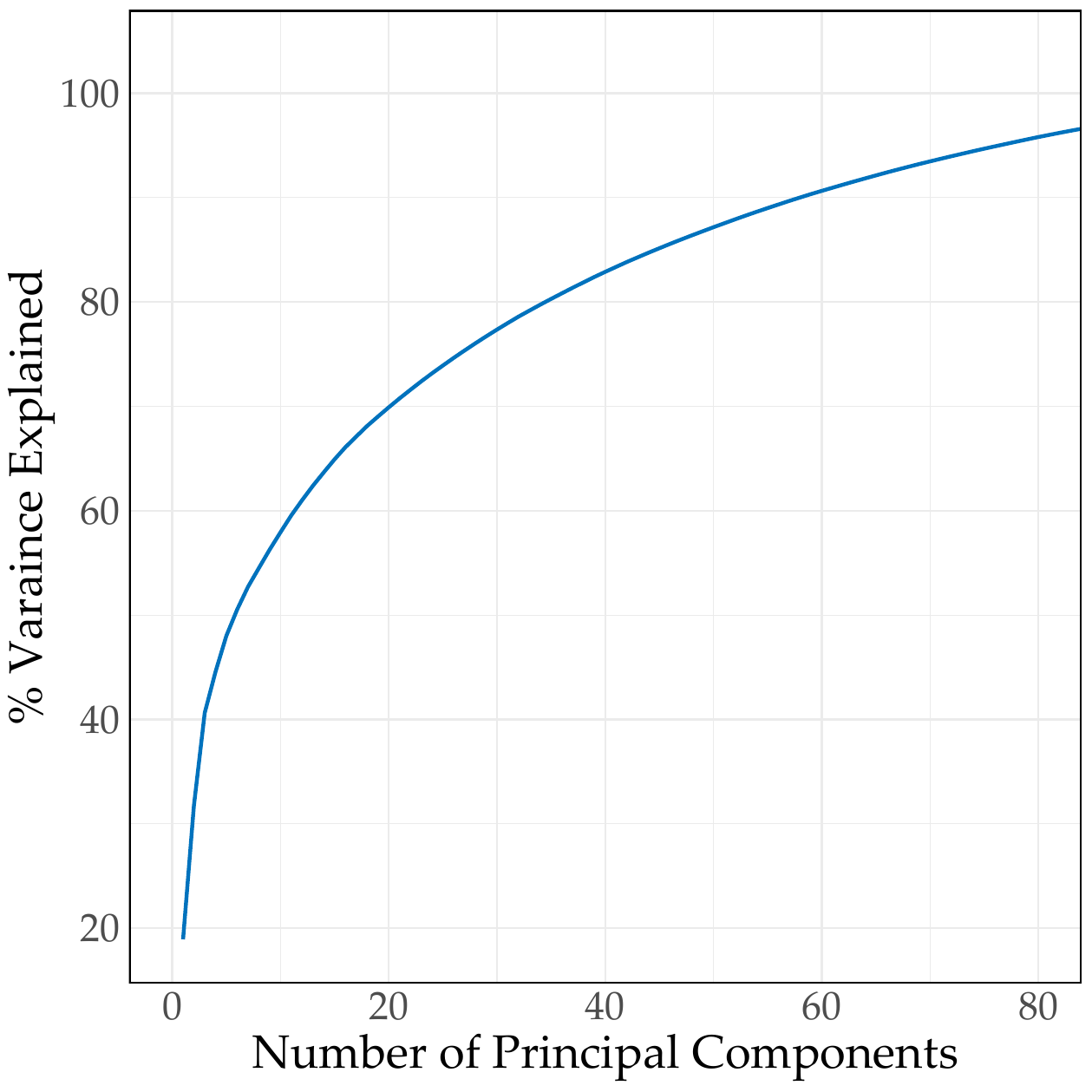}}
      \qquad
                               
\end{figure}

Figure \ref{fig:cor} illustrates this diversity.  It shows the distribution of correlations between pairs of long-short returns in the CZ22 data.  All long-short returns are signed to have positive mean returns, so highly negative correlations indicate very distinct predictors. 

The mean, median, and modal correlation is close to zero.  Almost all correlations fall below 0.5, indicating a zoo of distinct strategies.    This diversity is also seen in principal component analysis.  Panel (b) shows that it takes 60 principal components to span 90\% of total variance.  These weak correlations have been documented in many ways and in many papers (\citet{green2013supraview}; \citet{Mclean2016Does}; \citet{chen2020publication}; \citet{bessembinder2021time}; \citet{ChenZimmermann2021}).  

Weak correlations are intuitive given the publication process.  Referees at top tier finance journals are notoriously skeptical.  Moreover, they seem to believe in a strong factor structure in stock returns.  It's only natural that the predictors that make it through the referee process are distinct from other predictors.  

Weak correlations are also important for multiple testing statistics.  They imply that the small publication bias corrections found in \citet{harvey2016and} and \citet{chen2020publication} are valid.  We explain these estimates next.

\section{Just How Large Are  Publication Bias Effects in the Cross-Section?}\label{sec:gap}

\citet{Mclean2016Does}'s out-of-sample test (Fact \# 2, Section \ref{sec:facts-oos}) is a critical starting point.  It means publication bias drives at  \emph{most} 26\% of the returns reported in journals.   

However, the out-of-sample periods are  about  four years long, leading to standard errors of around 10 percentage points.  So loosely speaking, out-of-sample decay  only tells us that publication bias accounts for between 0\% and 46\% of published returns.

Using the gap between the empirical and null t-stat distribution (Fact \# 3) leads to more precise estimates.   Shrinkage and false discovery rate (FDR) methods transform this gap into estimates of publication bias effects, which are valid under weak dependence (Fact \# 4).  Three independent meta-studies find estimates that imply  publication bias effects account for 10\% to 15\% of in-sample performance  (\citet{harvey2016and}, \citet{chen2020publication}, \citet{Jensen2022Is}).

\subsection{Publication Bias Corrections}\label{sec:gap-mod}
Publication bias corrections can be understood in a four equation model.

Authors generates ideas for predictors. The t-stat for idea $i$ has two components:
\begin{align}\label{eq:mod-t}
    t_i &= \theta_i + Z_i \\ \label{eq:mod-Z}
    Z_i &\sim \text{Normal}(0,1) 
\end{align}
where $\theta_i$ is the  expected return divided by the standard error and $Z_i$ is sampling error divided by the standard error.  

$\theta_i$ is the key variable and has two interpretations.  The first is that it's the expected return (in units of standard errors).  The second is that it's a corrected t-statistic, since it corrects $t_i$ for sampling error $Z_i$, and thus publication bias.  We'll use both interpretations in what follows.

Since there are many ideas,  $\theta_i$ is not a constant, but  drawn from a distribution $f_\theta(\cdot|\sigma_\theta)$ 
\begin{align}\label{eq:mod-theta}
    \theta_i &\sim f_\theta(\cdot| \sigma_\theta)    
\end{align}
where $\sigma_\theta$ is a parameter vector that controls the properties of the distribution (e.g. variance).  We'll see this distribution is critical for determining publication bias corrections.

Publication bias means that observed ideas are a selected sample, given by
\begin{align}\label{eq:mod-pr-pub}
  \Pr(\pub_i | t_i, \theta_i)  &= p(t_i, \theta_i| \sigma_\pub)
\end{align}
where $\pub_i$ indicates observation and $\sigma_\pub$ is a parameter vector.  In general, $p(t_i,\theta_i | \sigma_\pub)$ is weakly increasing in both the t-stat $t_i$ and the underlying expected return  $\theta_i$.

Equations \eqref{eq:mod-t}-\eqref{eq:mod-pr-pub} typically imply that published t-stats overstate expected returns:
\begin{align}\label{eq:mod-essence}
     \E(t_i|\pub_i; \sigma_\theta, \sigma_\pub) 
     &= \E(\theta_i|\pub_i; \sigma_\theta, \sigma_\pub) 
     + \underbrace{\E(Z_i|\pub_i; \sigma_\theta, \sigma_\pub)}_{>0}.
\end{align}
This is the essence of publication bias.  If a paper says that the sample mean return is, say, three standard errors from zero ($t_i = 3$), the expected return is closer to zero ($\theta_i < 3$) because positive sampling errors are selected ($\E(Z_i|\pub_i; \sigma_\theta, \sigma_\pub)>0$).  Equation \eqref{eq:mod-essence} is very close to Equation \eqref{eq:intro}, but it makes explicit the fact that many tests have been run.  The many tests are embodied by $\sigma_\theta$, which contains information on an infinite number of  tests (Equation \eqref{eq:mod-theta}).

To correct for this bias, we just need to remove $\E(Z_i|\pub_i; \sigma_\theta, \sigma_\pub)$.  If we scale this term by the mean published t-stat, we get a ``shrinkage correction'':
\begin{align}\label{eq:mod-shrink}
    \text{Shrinkage}_\pub  \equiv 
    \frac{\E(Z_i | \pub_i; \sigma_\theta, \sigma_\pub)}
      {\E(t_i | \pub_i; \sigma_\theta, \sigma_\pub)},
\end{align} 
This says we have to shrink the mean published t-stat toward zero at a rate of $\text{Shrinkage}_\pub$ to recover the corresponding expected return.

Shrinkage measures the average correction.  The FDR measures  tail risk:
\begin{align}\label{eq:mod-fdrpub}
    \FDR_\pub = \Pr(\theta_i \le 0 | \pub_i; \sigma_\theta, \sigma_\pub)
\end{align}
That is, if we define ideas with zero or negative expected returns as ``false predictors'' and published predictors as ``discoveries,''  then the FDR is the risk of discovering a false predictor.  

There are many variations of shrinkage and the FDR.  Instead of conditioning on publication, these objects can condition on some other subset of tests (e.g. $ 2 < t_i  \le 3$).  The FDR can also be defined using $\theta_i = 0$ instead of $\theta_i \le 0$, or using the FDR definitions from \citet{benjamini1995controlling} or \citet{storey2003positive}. With many predictors and a non-trivial share of true predictors, these differences are more technical than meaningful (\citet{benjamini2008comment}). In the end, shrinkage measures the average correction and the FDR measures tail risk.

Using Equation \eqref{eq:mod-essence} to define corrections for publication bias  is explicitly described in \citet{chen2020publication} and is implicitly used throughout the meta-study literature on the cross-section of returns (\citet{Mclean2016Does}, \citet{Jensen2022Is}; \citet{harvey2016and}).\footnote{These publication bias corrections are different than \citet{andrews2019identification}'s Proposition 1, which extracts the distribution of $t_i$, rather than the distribution of $\theta_i | \pub_i$. }  This notion of a false discovery is consistent with the statistics literature (\citet{benjamini1995controlling};  \citet{storey2003positive}), but the recent finance literature uses looser language (Section \ref{sec:mis}).

\subsection{Publication Bias Corrections Are Roughly 10\% of In-Sample Returns}\label{sec:gap-ez}
The simplest estimates come from assuming normal (standardized) expected returns that are orthogonal to sampling error
\begin{align}\label{eq:ez-theta-norm}
    \theta_i &\sim \text{Normal}(0,\sigma_\theta^2) \\
    \Cov(\theta_i, Z_i) &= 0
\end{align}
and a clean hurdle for publication
\begin{align}\label{eq:ez-pr-pub-1}
  \Pr(\pub_i | t_i, \theta_i) = \begin{cases}
     \bar{p}, \quad t_i > 2   \\
     0, \quad \text{otherwise}.
  \end{cases}
\end{align}
This model is a slight generalization of  Appendix A.2 of \citet{chen2020publication}.

The model is also conservative.  Equation \eqref{eq:ez-theta-norm} assumes, pessimistically, that the ideas of finance scholars have  expected returns that are on average zero.  In the real world,  the intuition and  theory of finance scholars should  on average produce positive expected returns.  Similarly, Equation \eqref{eq:ez-pr-pub-1} assumes  $\pub_i$ is only related to  $\theta_i$ through $t_i$---a kind of pure data-mining setting.  In the reality, publication should also be positively related to $\theta_i$ through  the underlying quality of the paper (\citet{card2020editors}).  

These assumptions imply that $t_i \sim \text{Normal}(0,\sigma_\theta^2 + 1)$ and that $t_i | \pub_i$ is   just $t_i$ truncated from below at 2.  $\bar{p}$ drops out in the distribution of $t_i | \pub_i$, so the model just has a single meaningful parameter, the variance of expected returns $\sigma_\theta^2$ (in units of standard errors).

\begin{figure}[h!]
    \caption{
     Expected returns fill the gap.  Bars are $t_i$ from the CZ22 dataset.  Lines are a simple model (Equations \eqref{eq:mod-t}-\eqref{eq:mod-Z} and \eqref{eq:ez-theta-norm}-\eqref{eq:ez-pr-pub-1}) where the key parameter is $\sigma_\theta^2 = \Var(\theta_i)$, the variance of expected returns (in units of standard errors). The null of no predictability $\sigma_\theta = 0$ is very far from the data.  $\sigma_\theta \approx 3$ is required to fill the gap, implying that  expected returns three standard errors from zero are common.
    }
    \label{fig:filling the gap}
    \vspace{.10in}
    \centering     
                                \includegraphics[width=6in]{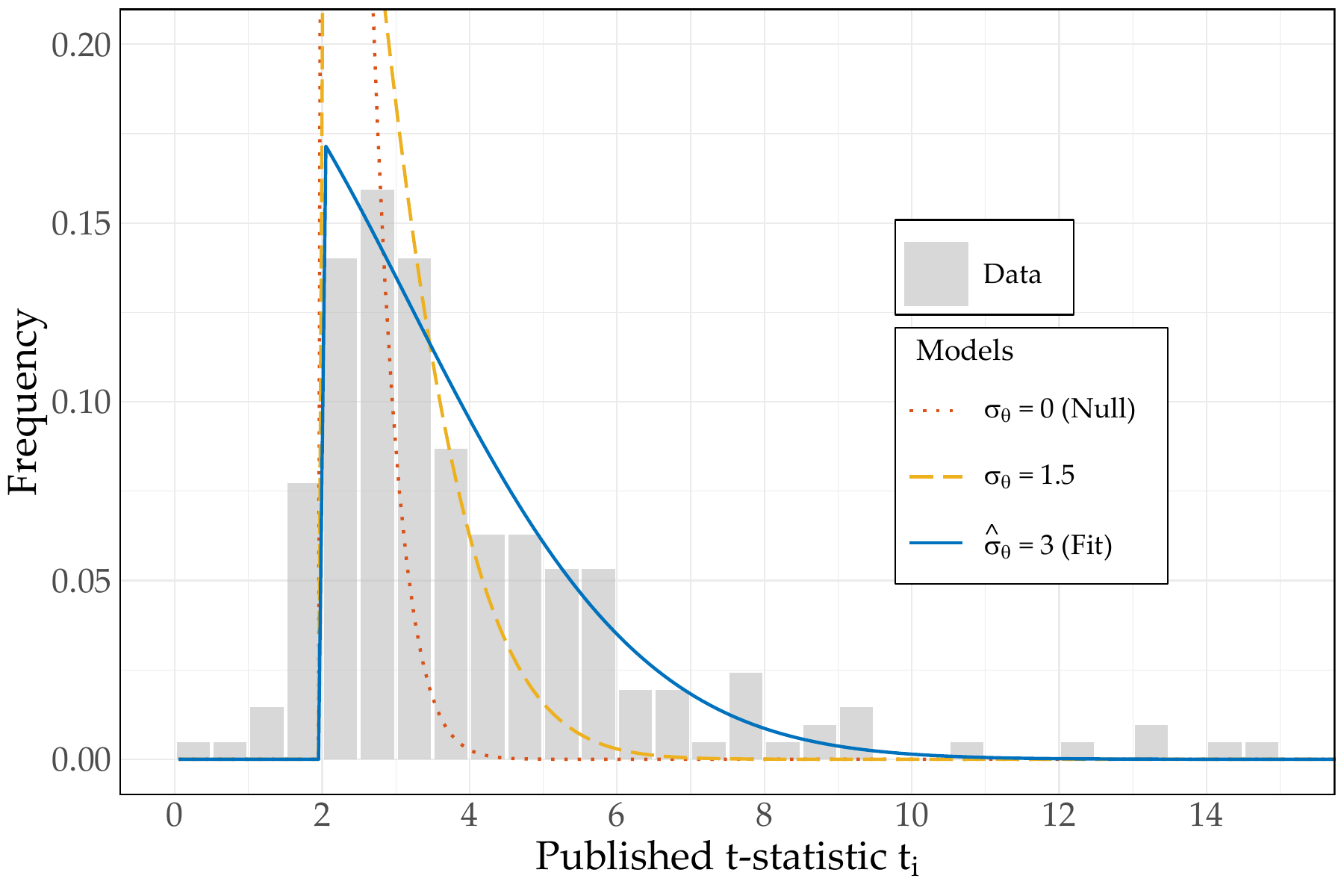}
                                                               
\end{figure}

$\sigma_\theta^2$ can be estimated by GMM or quasi maximum likelihood.  While $t_i$ and $t_j$ are not independent, Fact \#4 (Section \ref{sec:facts-cor})  shows the correlations are mild and center around zero, so standard econometric methods are valid (\citet{wooldridge1994estimation}).    Estimation is illustrated in Figure \ref{fig:filling the gap}.  For $\sigma_\theta^2 = 0$, the model (dotted line) is very far from the data.   Increasing $\sigma_\theta$ to 1.5 (dashed line) improves the fit, but the gap is still large.  $\sigma_\theta \approx 3.0$ (solid line) fills the gap, fitting many moments seen in this figure.  

What does $\hat{\sigma}_\theta \approx 3.0$ mean?  It means that finance researchers' ideas commonly have expected returns three standard errors away from zero.  Since the average standard error is about $20$ bps per month (\citet{chen2020publication}), this means that it's easy to find a predictor with a true expected return of $20\times 3 = 60$ bps per month.  It also means shrinkage and FDR estimates are small.

\begin{figure}[h!]
    \caption{Simple Shrinkage and FDR Estimates.  Plot compares the distribution of  t-stats (grey) and corrected t-stats (blue) conditional on publication, as implied by the simple model (Equations \eqref{eq:mod-t}-\eqref{eq:mod-Z} and \eqref{eq:ez-theta-norm}-\eqref{eq:ez-pr-pub-1}). The model uses $\hat{\sigma}_\theta^2 = \widehat{\Var}(\theta_i) = 3.0^2$ (Figure \ref{fig:filling the gap}).  Dashed lines show the means of each distribution.  The distance between these lines implies $\text{Shinkage}_\pub$ = 10\%.   $\FDR_\pub = 0.4\%$ is the tiny mass of corrected t-stats to the left of the solid line. }  
    \label{fig:monte-carlo-ez}
    \vspace{.10in}
    \centering     
    \includegraphics[width=0.8\textwidth]{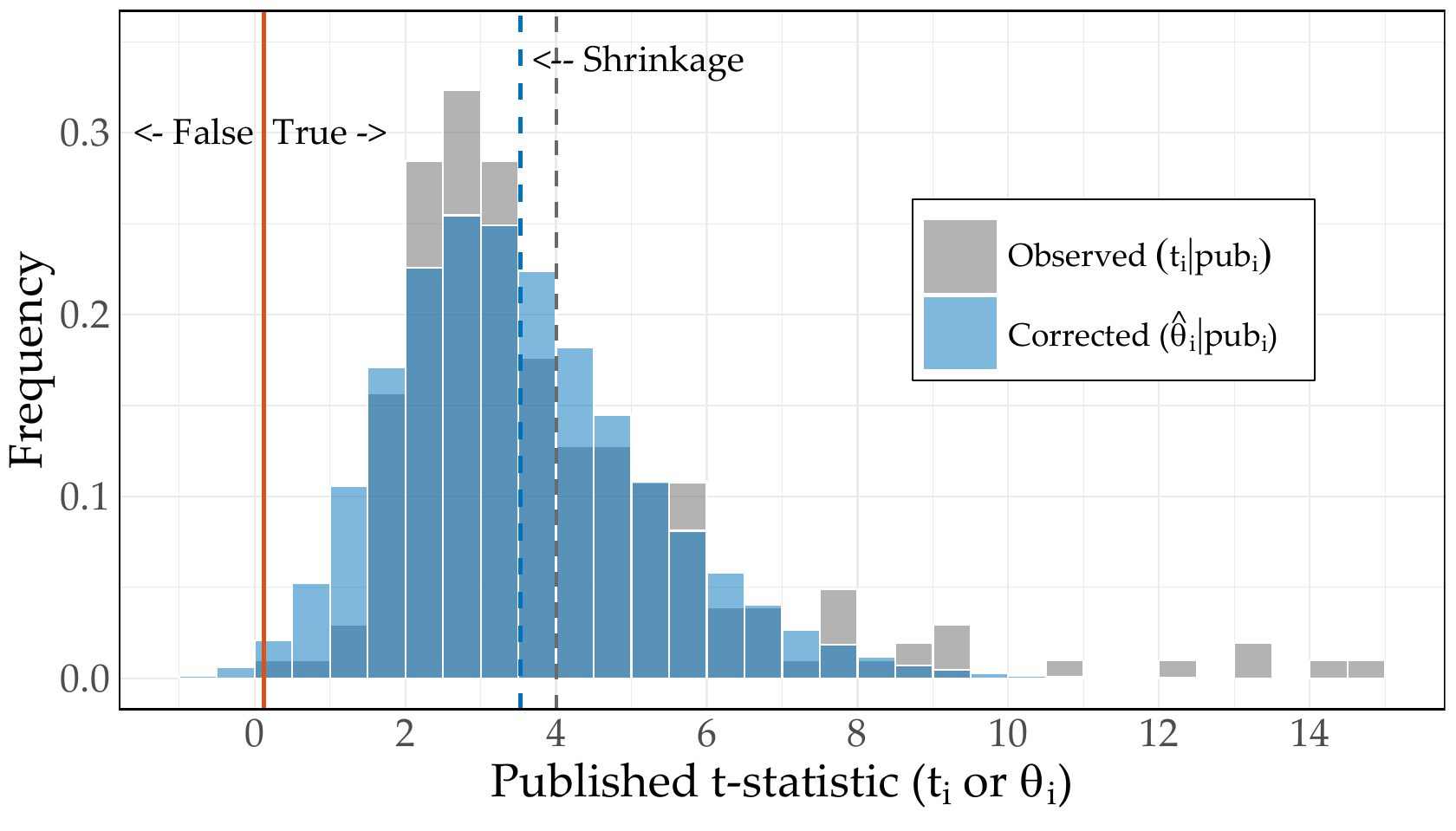}
\end{figure}

Figure \ref{fig:monte-carlo-ez} illustrates the implied shrinkage. We simulate $\theta_i$ and $t_i$ using  $\hat{\sigma}_\theta = 3.0$.  We then keep only published ideas using Equation \eqref{eq:ez-pr-pub-1}, and plot the distribution of $\theta_i | \pub_i$ (blue).  These standardized expected returns (alternatively, corrected t-stats)  are not far from the published t-stats (grey), showing that publication bias is small.  The small difference between the means of these distributions (dashed lines) shows that $\text{Shrinkage}_\pub$ is only 10\%.  

Bayesian reasoning provides the intuition:
\begin{align}\label{eq:ez-shrink-1}
    \E(\theta_i| t_i, \pub_i, \sigma_\theta) 
    &= 
    \E(\theta_i| t_i, \sigma_\theta)  
    = \left(1-\frac{1}
    {1 + \sigma_\theta^2}
    \right)t_i.
\end{align}
The first equality comes from assuming $\pub_i$ provides no information about $\theta_i$ over-and-above $t_i$ (Equation \eqref{eq:ez-pr-pub-1}), and is sometimes referred to as the ``immunity'' of Bayes rule to selection bias (\citet{efron2011tweedie}; \citet{chen2020publication}). The second equality uses standard bi-variate normal formulas.\footnote{To derive this, \begin{align}
 \E(\theta_i| t_i; \sigma_\theta) 
 &= \frac{\Cov\left(\theta_{i},t_{i} | \sigma_\theta \right)
    }{
    \Var\left(t_{i} | \sigma_\theta\right)
    }t_{i} 
    =\frac{\Cov\left(t_i - Z_i,t_{i} | \sigma_\theta\right)
    }{
    \Var\left(t_{i} | \sigma_\theta\right)
    }t_{i}
    = 
    1 - \frac{\Cov\left(Z_i,t_{i} | \sigma_\theta\right)
    }{
    \Var\left(t_{i} | \sigma_\theta\right)
    }t_{i}
\end{align}
and then note $\Cov\left(Z_i,t_{i} | \sigma_\theta\right)=1$ and  $\Var\left(t_{i} | \sigma_\theta\right) = 1+\sigma_\theta^2$.
}  

Equation \eqref{eq:ez-shrink-1} says we should shrink t-stats toward our ``prior,''  embodied in $\sigma_\theta$.  A small $\sigma_\theta$ is a strong ``prior'' that the expected return is zero, in which case shrinkage is large.  These ``priors'' are not  subjective beliefs of Bayesian statistics, but parameters estimated from empirical data.  For this reason, Equation \eqref{eq:ez-shrink-1} is often referred to as ``empirical Bayes'' shrinkage.  Plugging in our estimate of  $\hat{\sigma}_\theta = 3$,  we have 
\begin{align}\label{eq:ez-shrink-2}
      \E(\theta_i| t_i, \pub_i, \hat{\sigma}_\theta) 
    &= \left(1-\frac{1}{10}\right)t_i.
\end{align}
That is, published t-stats are overstated by just 10\%.

$\FDR_\pub$ is seen in the left tail of Figure \ref{fig:monte-carlo-ez}.  Only a tiny mass of the corrected t-stats fall to the left of the solid line, corresponding to  $\FDR_\pub = 0.4\%$.  The intuition  once again comes from Bayes rule:
\begin{align}\label{eq:ez-fdr-intuition}
    \FDR_\pub &= 
       \frac{\Pr(t_i>2 | \theta_i \le 0, \sigma_\theta)}{\Pr(t_i > 2 | \sigma_\theta)} 
       \Pr(\theta_i \le 0| \sigma_\theta) \\ \label{eq:ez-fdr-intuition-2}
       &\le 
       \frac{\Pr(t_i>2 | \theta_i = 0)}
       {\Pr(t_i > 2 | \sigma_\theta)} 
       \Pr(\theta_i \le 0)| \sigma_\theta).
\end{align}
where the second line comes from the fact that large t-stats are even less likely if $\theta_i$ is strictly less than zero.  This expression says we update our ``prior'' about $i$ being false ($\Pr(\theta_i \le 0)| \sigma_\theta)$) by comparing the null likelihood $\Pr(t_i>2 | \theta_i = 0)$ to the data $\Pr(t_i > 2 | \sigma_\theta)$.  If the null and data are far apart, this fraction is small, and $\FDR_\pub$ is small. Plugging in our estimated $\theta_i \sim \text{Normal}(0,3^2)$ and $t_i \sim  \text{Normal}(0, 10)$:
\begin{align}
    \FDR_\pub 
       &\le 
       \frac{0.025}{0.25} 
       (0.5)
       = 4.8\%.
\end{align}
That is, the risk of a published predictor having zero or negative returns is at most 4.8\%.  

\subsection{Publication Bias Estimates Implied by Meta-Studies Are Also Small}\label{sec:gap-lit}

\citet*{harvey2016and}; \citet{chen2020publication};  and \citet*{Jensen2022Is} estimate the  distribution of corrected t-stats ($\theta_i$) using methods that are much more sophisticated than those in Section \ref{sec:gap-ez}.  Nevertheless, the resulting distributions are quite similar: all find that expected returns three standard errors away from zero are common. 

\begin{figure}[h!]
    \caption{Distribution of Corrected t-stats  ($\theta_i$) from the Literature.   \citet*{harvey2016and} uses their baseline SMM estimate (their Table 5, Panel A, $\rho = 0.2$).  \citet{chen2020publication} uses their baseline (Table 3, ``All'').  \citet*{Jensen2022Is} uses their baseline publication bias adjustment (Figure 9, $\tau_c = 0.29\%$).  ``Simple Normal'' uses Section \ref{sec:gap-ez} (based on \cite{chen2020publication}'s appendix).  The literature differs in the modeling of the null ($\theta_i \le 0$) but for $\theta_i > 0$ the distributions are similar to the simple normal model (Figure \ref{fig:filling the gap}).  All find that expected returns three standard errors from zero are common.
    }
    \label{fig:lit-comp}
    \vspace{.10in}
    \centering     
\includegraphics[width=0.8\textwidth]{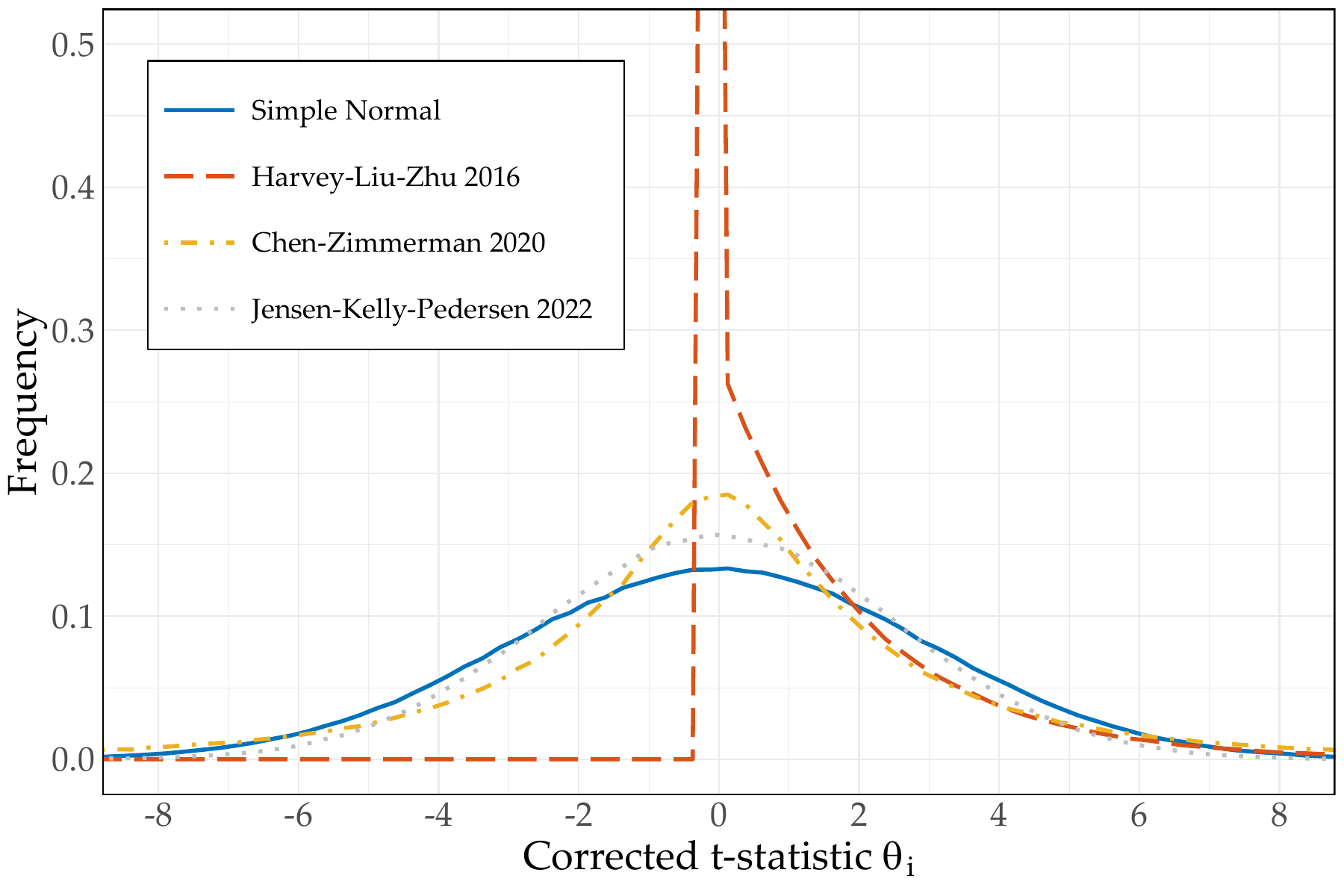}
\end{figure}

The distributions of $\theta_i$ from these papers are shown in Figure \ref{fig:lit-comp}.  These papers differ in their modeling of null predictors: \citet{harvey2016and} put all null predictors into the mass near zero, while \citet{chen2020publication} and \citet{Jensen2022Is} allow for negative $\theta_i$.  This results in differences for $\theta_i \le 0$, but to the right of zero the distributions are quite similar.  They all imply that predictors with $\theta_i > 0$ on average have $\theta_i \approx 2$, not far from our one-parameter estimate.

This commonality is striking because  the estimates are made by  independent teams. The teams differ in members, data sources, publication modeling, expected return modeling, and estimation method.  Some t-stats were hand-collected, while others were replicated.   Some modeled Equation \eqref{eq:mod-pr-pub} explicitly, while others used informal adjustments.  Expected returns were modeled as mixtures, t-distributed, or hierarchical normal. Even the estimation differed, from SMM to (quasi) maximum likelihood to full maximum likelihood.  All of these different methods lead to similar inferences about the right tail of $\theta_i$.  Comparable estimates are also found in \citet*{chinco2021estimating}, using yet another approach.

The common  tail leads to similar shrinkage  and $\FDR_\pub$ estimates. \citet{chen2020publication} reports $\text{Shrinkage}_\pub$ estimates between 8\% and 17\% across a wide range of models.  In the working paper version (\citet{chen2018publication}) we also report $\FDR_\pub \approx 1\%$.  \citet{Jensen2022Is} report shrinkage of 10\% and an $\FDR_\pub \approx 0.1\%$ before adjusting for publication bias, but  adjusting for publication bias has only minor effects.   A simple calculation shows their publication bias adjusted shrinkage is around 13\%.\footnote{\citet{Jensen2022Is}'s footnote 36 states shrinkage is approximately \begin{align}
    1-\frac{1}{1+(0.10^2/12)/(420 \tau^2)}
\end{align}
where $\tau^2 = \tau_c^2 + \tau_w^2$.  Plugging in the publication bias adjusted $\tau_c = 0.29\%$ and $\tau_w=0.21\%$ leads to shrinkage of 13\%.}

\begin{figure}[h!]
    \caption{Shrinkage and FDR Estimates in \citet{harvey2016and}.  
     Plot compares the distribution of published t-stats (grey) to the distribution of standardized expected returns (blue) implied by \citet{harvey2016and}'s baseline estimate  (Table 5, Panel A, $\rho = 0.2$).  Dashed lines show the means of each distribution.  The distance between these lines implies $\text{Shrinkage}_\pub$ = 13\%.   $\FDR_\pub = 6.3\%$ is the mass of expected returns to the left of the solid line. These corrections are similar to the simple model (Figure \ref{fig:monte-carlo-ez}) because the right tail of $\theta_i$ is similar (Figure \ref{fig:lit-comp}).  
    }
    \label{fig:monte-carlo-hlz}
    \vspace{.10in}
    \centering     
    \includegraphics[width=0.8\textwidth]{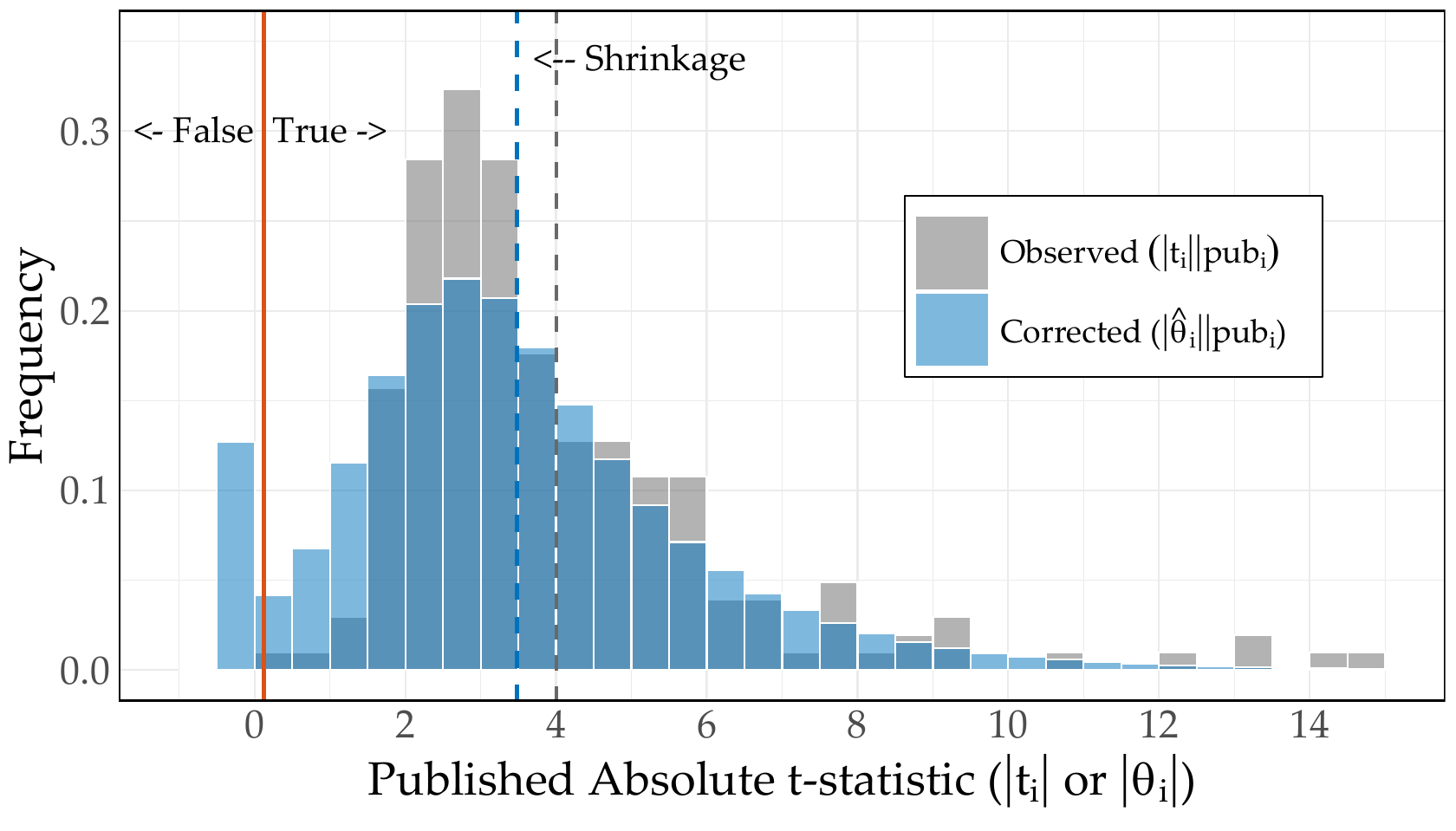}
\end{figure}

\citet{harvey2016and} don't report shrinkage or $\FDR_\pub$, but these corrections can be easily calculated by Monte Carlo.  Figure \ref{fig:monte-carlo-hlz} illustrates the calculation.  We simulate \citet{harvey2016and}'s baseline model and plot the distribution of published corrected t-stats $\theta_i|\pub_i$. A minor complication is that \citet{harvey2016and} assume publication depends on $|t_i|$ rather than $t_i$.\footnote{This is not explicitly stated in \citet{harvey2016and}, but we find this assumption is required to replicate their Table 5.}  Nevertheless, the corrected t-stats (blue) are close to their observed counterparts $|t_i||\pub_i$ (grey).   $\text{Shrinkage}_\pub = 13\%$ and  $\FDR_\pub = 6.3\%$.  

\subsection{Model-Free Supporting Evidence}\label{sec:modfree}

\citet{harvey2016and}, \citet{chen2020publication}, and \citet{Jensen2022Is} use models to extrapolate the distribution of $\theta_i$ from published results.  These models all lead to similar bias corrections, but one would like to see model-free evidence.

Model free evidence comes from the data-mining experiments of \citet{yan2017fundamental} and \citet*{Chordia2020Anomalies}.  These papers systematically mine Compustat accounting data for signals to trade on.  Compustat contains more than 200 accounting variables, and since signals typically combine at least two variables, it's quite easy to generate 200 choose 2 $\approx$ 20,000 signals.  Yan and Zheng generate 18,000 signals, while Chordia et al. generate 2 million.  By comparing the t-stats from these signals to the null distribution, one can get a sense of the distribution of $\theta_i$ from pure data mining.

The distribution of $|t_i|$ found in the \citet{yan2017fundamental} data is shown in Table \ref{tab:t-too-big} (back in Section \ref{sec:facts-gap}).  The distribution is very far from the null.  30\% of $|t_i|$ exceed 2.0, six times the null probability of 5\%.  The distance becomes extremely pronounced further in the tail.  4.5934\% of $|t_i|$ exceed 5.0,  20,000 times more common than the null probability of 0.0002\%.   The distribution found in \citet{Chordia2020Anomalies} is not quite so dispersed because they focus on value-weighting.  Nevertheless, \citet{Chordia2020Anomalies} find that 10 to 20 percent of $|t_i|$ exceed 2.0, at least double the 5\% found under the null.  Neither paper finds a huge spike in t-stats near zero, suggesting that the smooth modeling of $\theta_i$ in \citet{chen2020publication} and \citet{Jensen2022Is} is more appropriate than the Dirac delta mixture in \citet{harvey2016and} (Figure \ref{fig:lit-comp}).




\section{Interpreting and Mis-Interpreting Meta-Studies}\label{sec:mis}

Facts 1-3 (Section \ref{sec:facts}) are straightforward to interpret.  Each fact measures either the author error or the sampling error in published studies.   Both types of error are small, even if data-mining is widespread (Section \ref{sec:gap}).

Other facts from meta-studies are more tricky.  Returns are about 50\% weaker far from the original samples. Returns are about 30\% weaker after liquidity adjustments.  Many predictors are insignificant under some multiple testing methods.

It tempting to interpret these negative results as evidence of publication bias (i.e. data snooping, p-hacking, etc).  But since Facts 1-3 limit the role of author error and sampling error,  these findings of return decay largely document variation in expected returns (Equation \eqref{eq:intro}).  This variation is important, but it does not measure publication bias effects.  

The finding that many predictors are insignificant in multiple testing methods is not like the others.  It is fragile and depends entirely on  highly conservative hypothesis tests. 

\subsection{Many Predictors Are Insignificant Under Some Many-Hypothesis Methods}\label{sec:mis-hurdle}

\begin{figure}[h!]
    \caption{Multiple Testing vs Conservatism in  \citet*{harvey2016and}.  We simulate predictors following
    \citet{harvey2016and}'s baseline estimate (Table 5, Panel A, $\rho = 0.2$).   $\text{Normal}(0,0.1)$ noise is added to the false predictors for ease of viewing. We plot a random sample of 1,172 total predictors, which implies roughly 300 predictors with $|t_i|>1.96$.     The classical 1.96 hurdle implies an FDR of 8.8\% (share of hollow markers to the right of solid line).  FDR = 5\% requires raising the hurdle a bit, to 2.3 (dashed line).  \citet{harvey2016and} recommend a more a conservative FDR = 1\% (dotted line), or using the \citet{holm1979simple} algorithm at FWER $\le$ 5\% (dotted line), among other more conservative methods.  Holm  is computed using the 1,383 predictors shown.  The significant raising of hurdles is due to conservatism, not multiple testing effects.
    }
    \label{fig:monte-carlo-hlz-scatter}
    \vspace{.10in}
    \centering     
    \includegraphics[width=0.8\textwidth]{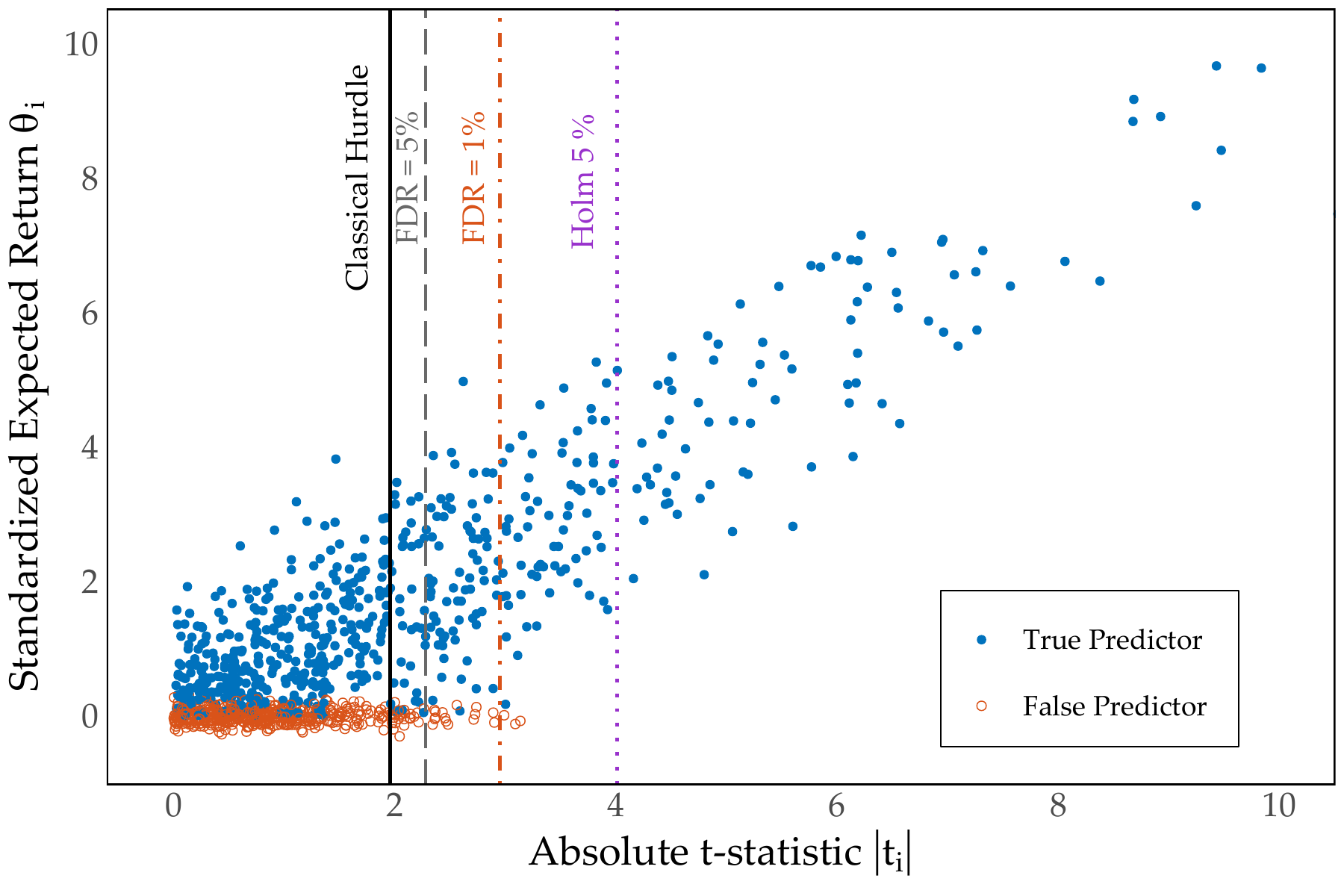}
\end{figure}

Figure \ref{fig:monte-carlo-hlz-scatter} illustrates this finding by simulating \citet{harvey2016and}'s baseline estimate.  It plots the standardized expected return $\theta_i$ against the absolute t-stat $|t_i|$, with each marker representing one predictor.  \citet{harvey2016and}'s model assumes $\theta_i = 0$ for all false predictors, which makes them hard to see in a scatterplot.  We add $\text{Normal}(0,0.1)$ noise to these markers  for ease of viewing. 

The classical hurdle of 1.96 implies an FDR of only 8.8\%, as seen in the small share of false predictors (empty markers) to the right of $|t_i| = 1.96$.  To achieve an FDR of 5\%, the hurdle should be raised slightly, to 2.3 (dashed line).   Given the uncertainty in these estimates, it is reasonable to think that the classical hurdle need not be raised at all. One would like to examine standard errors here, but \citet{harvey2016and} do not report them.\footnote{In ongoing work, \citet{chen2022t} finds that the standard error on these hurdles is approximately 1.0.}

FDR = 5\% is not only a natural counterpart to the \citet{fisher1925statistical}'s 5\%.  It is also a  popular choice in fields from genomics to functional imaging (\citet{benjamini2020selective}).   But \citet{harvey2016and} focus on FDR = 1\%, five times more stringent than the traditional level.  This choice calls for  raising the hurdle substantially, to 3.0 (dash-dot line), leading to many insignificant findings to the right of 1.96.  However, 81\% of these insignificant findings are true predictors and have an average expected return 1.6 standard errors from zero.  These insignificant findings say little about publication bias.  They can only be interpreted as increased conservatism. 

Conservatism  drives other calls for raising the classical hurdle.  \citet{harvey2016and} also justify raising the hurdle using Theorem 1.3 of  \citet{benjamini2001control}.  This method is thought by the statistics community to be excessively conservative.  For example, in his textbook on large scale hypothesis testing, \citet{efron2012large} describes Theorem 1.3 of Benjamini-Yekutieli (2001) as a ``severe penalty'' and ``not really necessary.''\footnote{ Even the original \citet{benjamini2001control} describes the algorithm as ``very often unneeded, and yields too conservative a procedure.'' The bulk of that paper supports the original, less conservative \citet{benjamini1995controlling} algorithm (Theorem 1.2).  This original algorithm is shown to work under weak dependence (\citet{storey2004strong}; \citet{ferreira2006benjamini}; \citet{farcomeni2007some}; \citet{genovese2006false}).  In his lecture notes ``A Tutorial on False Discovery Control,'' Chris Genovese says that the original algorithm is ``quite hard to break even beyond what has been proven.''  Unlike in finance, the ``BY Algorithm'' in statistics typically refers to a different paper, \citet{benjamini2005false}, which is based on the \citet{benjamini1995controlling} algorithm.
} \citet{harvey2016and} also examine the  Bonferroni and Holm methods, specifying a family wise error rate (FWER) $\le 5\%$ (Figure \ref{fig:monte-carlo-hlz-scatter}, purple line).  These tests are  also extremely conservative.  They imply that \emph{every single} significant predictor is true---with 95\% probability (dotted line, Figure \ref{fig:monte-carlo-hlz-scatter}). \citet{Chordia2020Anomalies} recommend a hurdle of 4.0 using the \citet{romano2008formalized} method.  Their hurdle implies that the probability the false discovery proportion exceeds 5\% is less than 5\%.   It  basically limits the tail risk of a tail risk to only 5\%.   


\subsection{Predictability is 50\% Weaker Far From the Original Sample Periods}\label{sec:mis-oos}

\citet{Mclean2016Does} document that post-publication returns are 50\%  smaller than in-sample returns.  This fact has been replicated many times (\citet{Jacobs2020Anomalies}; \citet{chen2020publication}; \citet{Jensen2022Is}) and is known as ``anomaly decay.''

\begin{figure}[h!]
    \caption{
    Predictability is 50\% Weaker After Publication.  Returns are normalized so that the mean in-sample return is 100 bps per month (blue line).  Returns are averaged  across all CZ22 predictors  within each month, then the trailing 36-month mean is computed.   The mean post-publication return is in red.  Publication bias effects are bounded by decay  three years after publication (dashed line).  Shrinkage estimates imply half of this decay is publication bias (yellow line).  The remaining decay is due to changes in expected returns, which are natural given the claim in many of the original papers that predictability is due to mispricing (e.g. \citet{rosenberg1985persuasive}). 
    }
    \label{fig:roll-rbar}
    \vspace{.10in}
    \centering     
    \includegraphics[width=0.8\textwidth]{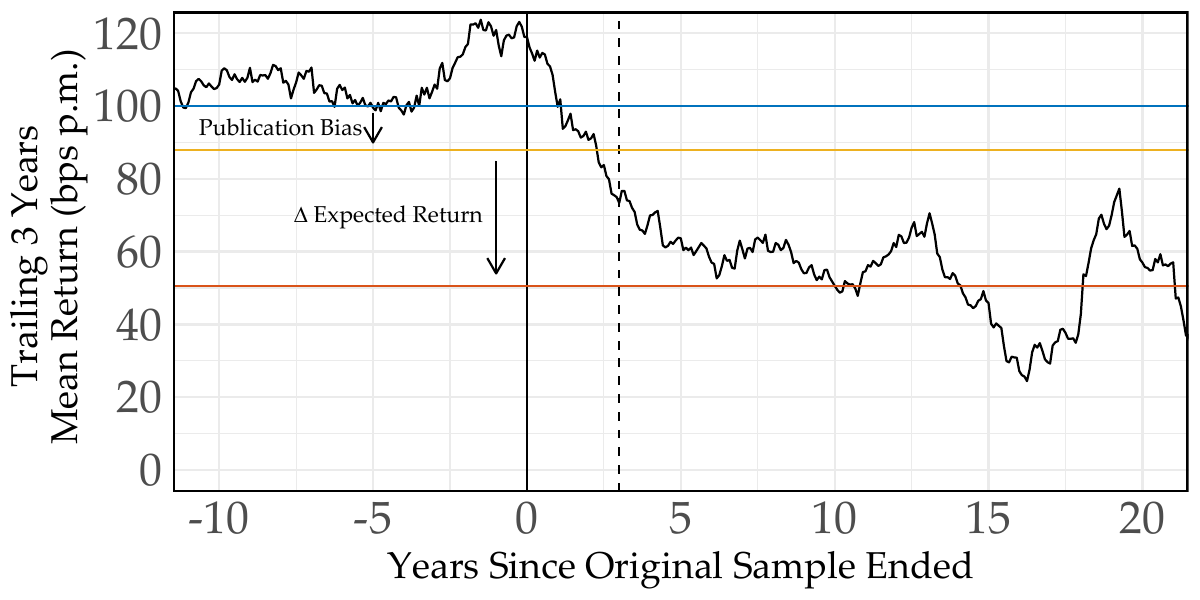}       
\end{figure}

Figure \ref{fig:roll-rbar}  visualizes this decline by plotting trailing 3-year mean returns in event time, where the event is the end of the original sample period.  This trailing mean declines smoothly after the sample ends.  Returns decline by about 25\% three years post-sample, and drop by 40\% five years post-sample.  10 years after the original samples end, the trailing 3-year mean reaches the 50\% decline found in the full post-publication sample (red line).

This fact is often mis-interpreted as evidence of pervasive publication bias effects (data snooping, p-hacking, etc).  The problem with this interpretation is that monthly predictor returns have very little serial autocorrelation.  The mean autocorrelation is 7\% at 1 month and falls to zero at 3 months.  So return decay driven by publication bias should show up just months after the original samples end, not a decade later.    More precisely, publication bias effects are bounded by the 25\% decline at year 3 (Fact \#2, dashed vertical line).  

Multiple testing statistics provide a  direct estimate of publication bias effects (Fact \# 3, Section \ref{sec:gap}) and imply a roughly 12\% decline (yellow line).  The remaining 38\% decline, then, is a decline in the  expected return.  This decline is important for interpreting the economic meaning of predictability, but it is not publication bias.

The economic meaning becomes clear if one reads the original papers.  Many, if not most  papers argue that their finding is due to mispricing (\citet{rosenberg1985persuasive}; \citet{jegadeesh1993returns};  etc).  Other papers simply present their findings as a puzzle (e.g. \citet{banz1981relationship}; \citet*{foster1984earnings}), which is consistent with the mispricing explanations.  In contrast, claiming that  mispricing is permanent or that the empirical relationship is somehow structural is less common.   It should not be surprising if mispricing is traded away after it is publicized (\citet{marquering2006disappearing}) or if improvements in liquidity decrease mispricing (\citet{Chordia2014Have}).  There's a reason why this literature is called ``anomalies.''

Returns are also about 50\% smaller in non-U.S. markets (\citet{Jacobs2020Anomalies}; \citet{Jensen2022Is}) and decline similarly in long samples before the original sample periods begin (\citet{Linnainmaa2018The}).  These results further illustrate how anomalies depend on the investing environment.  Predictability may rely on the uninformed trades of retail investors (\citet{goetzmann2018momentum})  or small, illiquid stocks (\citet{novy2015taxonomy}), both of  which are less prevalent outside the U.S. (\citet{ke2018cross}), and before the original samples begin (typically before 1963).  Like post-publication decay, these findings are important for understanding the nature of predictability, but they do not measure publication bias, precisely defined.

\subsection{Predictability is about 30\% Weaker After Liquidity Adjustments}\label{sec:mis-liq}

Finance researchers have long known that predictability is weaker if rebalancing is less frequent (\citet{jegadeesh1993returns}) or if trading is confined to large, liquid stocks (\citet{fama1993common}; \citet{ball1995problems}).  This result has a natural interpretation of liquidity effects (\citet{mayshar1979transaction}; \citet{kyle1985continuous}) and is closely linked to the idea of costly arbitrage (\citet{pontiff1996costly}; \citet{shleifer1997limits}).  In strategies that avoid illiquidity, arbitrage is more likely, mispricing is smaller, and predictability is weaker.

\begin{figure}[h!]
    \caption{
    Liquidity Adjustments Decrease Returns by 30\%.  Data are 207 predictors from CZ22.  Grand mean return averages across in-sample months and then averages across predictors.  Error bars show to standard errors, approximated by the standard deviation across predictors divided by $\sqrt{207}$.   Original implementations follows the original papers.  Annual rebalancing updates signal data each year in June.  ME $>$ NYSE 20 Pct excludes stocks that fall below the 20th percentile of NYSE market equity.  Value-weighted weights stocks by their market equity.  Liquidity adjustments robustly decrease expected returns by roughly 30\%.  Robust effects related to economics should not be equated with data snooping.
    }
    \label{fig:liq}
    \vspace{.10in}
    \centering     
    \includegraphics[width=4in]{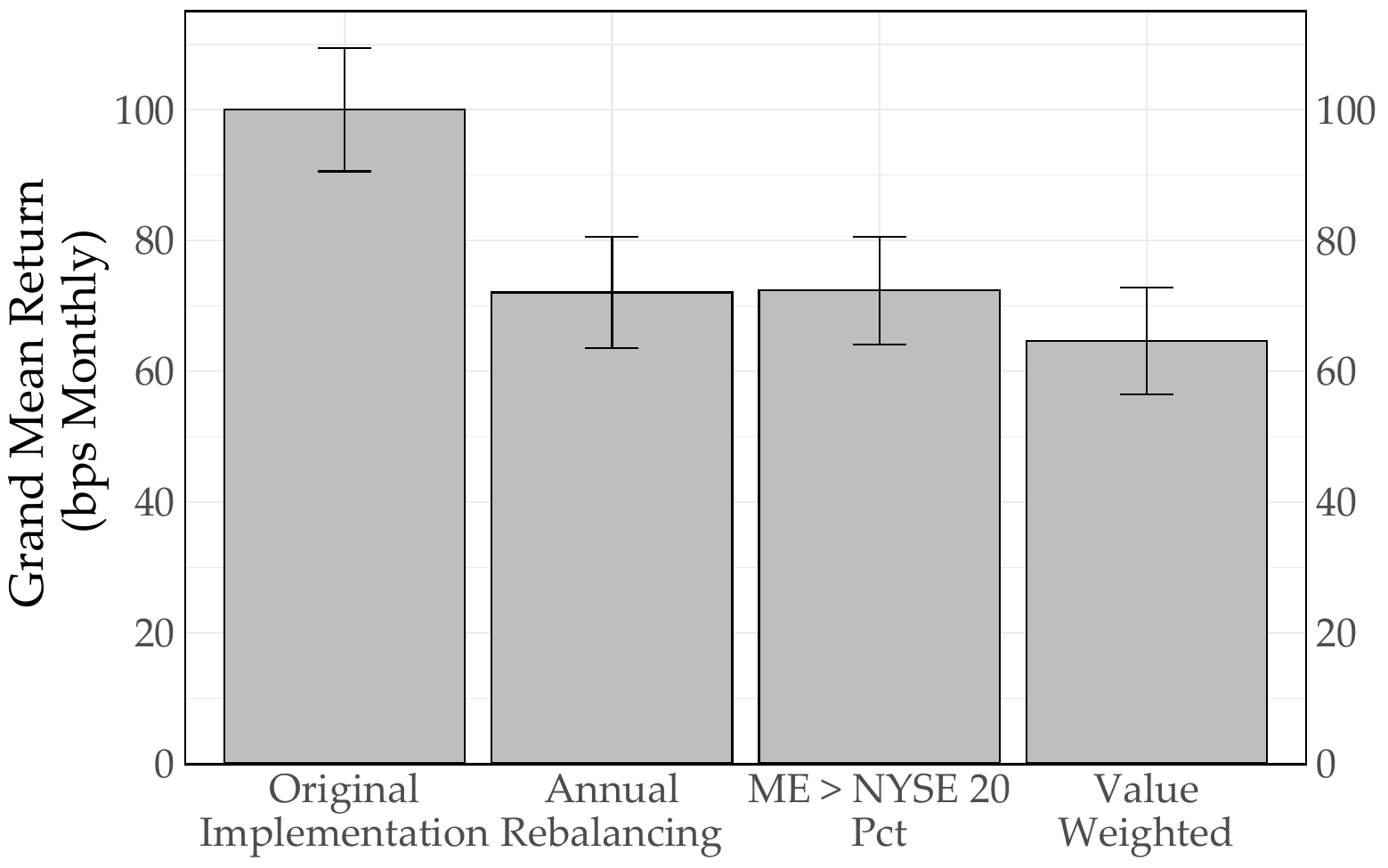}       
\end{figure}

These early findings turn out to be representative of the broader cross-sectional predictability literature (\citet{novy2015taxonomy}; \citet{Jacobs2020Anomalies}; \citet{chen2022zeroing}). Figure \ref{fig:liq} replicates these results using the CZ22 dataset.  The original implementations typically rebalance monthly and almost all equally weight stocks in each leg of the portfolio.  Rebalancing annually reduces returns by 25\%, as does restricting trading to stocks above the 20th percentile of NYSE market equity.  Value-weighting each leg of the portfolio reduces returns even more (roughly 35\%). 

Strangely, the recent meta-study literature sometimes cites ``data snooping,'' or ``p-hacking''  in connection with  Figure \ref{fig:liq}.   Rigorously defined, these terms refer to chance results that are unrelated to economic phenomena (e.g. \citet{Lo1990Data}; \citet{white2000reality}).  In contrast, the liquidity effects in Figure \ref{fig:liq} are  robust and founded in  classic finance theory.  

Liquidity effects are important.  One can argue that the original implementations in Figure \ref{fig:liq} are irrelevant, and that the liquidity adjustments better capture their potential returns.    On the other hand, one can argue that the original implementations provide more powerful tests of classical equilibrium theories, or that investors would  combine predictors anyway (\citet{DeMiguel2020transaction}).

These are all arguments about what is the proper ``true effect'' that one should be trying to measure in Equation \eqref{eq:intro}.  These debates require insight into portfolio choice, market equilibrium, and market microstructure, and go far outside the realm of publication bias.

\subsection{Loose Language in Meta-Studies}

The misinterpretations in Sections \ref{sec:mis-hurdle}-\ref{sec:mis-liq} are due to loose language.  Changes in expected returns should not be equated with data snooping or p-hacking.  Many insignificant predictors  do not imply multiple testing problems, especially if the significance tests are extremely conservative.    

Meta-studies are prone to loose language, as each study covers a lot of ground, making precise language difficult.  Meta-studies also use statistics unfamiliar to finance researchers.  

However, precise language is possible and important.   In everyday life, loose language is spoken and causes confusion.  In academic publications, loose language is written down, and can cause confusion for future generations of researchers.

We emphasize the precise usage of language by taking a closer look at two influential meta-studies, \citet{harvey2016and}  and \citet{Hou2020Replicating}.

\subsubsection{Harvey, Liu, and Zhu's (2016) ``False Findings''}
\citet*{harvey2016and} (HLZ) famously ``argue that most findings in financial economics are likely false.'' Yet their own estimates imply a false discovery rate of less than 10\%.  (Section \ref{sec:gap-lit}). So where does HLZ's famous quote come from?

Their conclusion suggests an answer:
\begin{quotation}
\emph{In medical research, the recognition of the multiple testing problem has
led to the disturbing conclusion that “most claimed research findings are
false” (Ioannidis (2005)). Our analysis of factor discoveries leads to the same conclusion–many of the factors discovered in the field of finance are likely false discoveries: of the 296 published significant factors, 158 would be considered false discoveries under Bonferonni [sic], 142 under Holm, 132 under [\citet{benjamini2001control} Theorem 1.3] (1\%),
and 80 under [\citet{benjamini2001control} Theorem 1.3] (5\%).}
\end{quotation}

This passage conflates ``most'' with ``many'' and  ``false discovery'' with ``insignificant predictor.''  142 out of 296 is not most.  And while Bonferroni, \citet{holm1979simple}, and \citet{benjamini2001control}'s Theorem 1.3 can determine the number of insignificant results, they cannot determine the number of false discoveries.  Using more precise language, HLZ's claim would be: ``many findings in cross-sectional asset pricing are insignificant under some highly conservative multiple testing adjustments''  (see Section \ref{sec:mis-hurdle})

``False'' may seem equivalent to ``insignificant,'' but these two terms have an important distinction in statistics.  In statistics, a ``false discovery'' or ``false rejection'' is a significant result that is, in truth, null  (\citet{soric1989statistical}; \citet{benjamini1995controlling}; \citet{storey2011false}).   This distinction is fundamental.  It is, in essence, ``Principal 2'' of  the American Statistical Association's ``Statement on Statistical Significance and P-Values'' (\citet{wasserstein2016asa}).\footnote{One may be tempted to define a ``false discovery'' as a ``significant finding under a classical test that would not have been considered significant under a many hypothesis algorithm.''  The problem with this definition is that there is no objective way to choose the many hypothesis algorithm.  Subjectivity is a problem even in classical significance tests (\citet{mcshane2019abandon}).}  

This paragraph also cites a theoretical opinion essay (\citet{ioannidis2005most}) as if it is the conclusion of medical research.  The empirical counterpart to Ioannidis finds a false discovery rate of only 14\% in top medical journals  (\citet{jager2014estimate}).  While this single study has its limitations,  review papers find other empirical studies at odds with Ioannidis's claims, and none in support (\citet{leek2017most}; \citet{fanelli2018science}).  

Unfortunately, loose language is echoed in follow-up papers.  In fact, HLZ's citation of  \citet{ioannidis2005most} is an example.  \citet{ioannidis2005most} claims ``[i]t can be proven that most claimed research findings are false.''  Yet his claim depends on parameter values that are not universally accepted (e.g., \citet{goodman2007most}; \citet{christensen2018transparency}). Thus, a more precise claim would be ``under certain parameter values, most claimed research findings are false.''  


Harvey et al.'s  mixing of ``insignificant'' and ``false'' is echoed in many papers, including \citet{Hou2020Replicating}; \citet{Chordia2020Anomalies}; and our own paper \citet{ChenZimmermann2021}.  We urge  finance researchers to use precise language and spare future generations of this confusion.

\subsubsection{Hou, Xue, and Zhang's (2020) ``Failed Replications'' and Microcaps}

The second paragraph of \citet{Hou2020Replicating} (HXZ) summarizes the paper:
\begin{quotation}
\emph{Our key finding is that most anomalies fail to replicate, falling short
of currently acceptable standards for empirical finance. First, of the 452
anomalies, 65\% cannot clear the single test hurdle of $|t| \ge 1.96$.  The key
word is ``microcaps'' Microcaps represent only 3.2\% of the aggregate market
capitalization but 60.7\% of the number of stocks. Microcaps have the highest
equal-weighted returns and the largest cross-sectional dispersions in returns
and in anomaly variables.}
\end{quotation}
One would assume that HXZ replicate 452 anomalies but that HXZ's tests fail, mostly because they downweight microcaps.  But this is not  the case.

Just a few paragraphs later, HXZ explain that microcaps have a minor role:
\begin{quotation}
\emph{The failure rate drops to 43.1\% if we allow microcaps to run amok with NYSE-Amex-NASDAQ breakpoints and equal-weighted returns.}
\end{quotation}
In other words, microcaps account for only $1-43.1/65=34\%$ of their failed replications.  

If it's not microcaps, then what accounts for the majority of the failed replications?   It's hard to tell because of loose language.   HXZ do not define what they consider an ``anomaly,'' and most of HXZ's 452 ``anomalies'' were never shown to obtain $|t| \ge 1.96$ in the original papers (\citet{ChenZimmermann2021}; \citet{Jensen2022Is}).  So many of the ``failed replications'' are actually successful replications.   For many ``anomalies,'' HXZ apply ``small perturbations to the original variable definitions.''  These perturbations are not done systematically, but applied ``when necessary.'' HXZ do not define when it is necessary, nor do they clearly define which perturbations should be applied when it is necessary.  To our reading, the perturbations and their applications can only be described as ad-hoc.\footnote{For example, HXZ list Dichev's Z-score, \citet{francis2004costs}'s many earnings attributes, and \citet{campbell2008search}'s failure probability as ``influential anomalies that failed to replicate.'' But \emph{none} of these variables was shown to have statistically significant predictive power using raw long-short returns in the original papers. HXZ found it necessary to perturb the Z-Score and failure probability, making eight ``anomalies'' out of these two variables.  But they did not find it necessary to perturb \citet{francis2004costs}'s earnings persistence, earnings predictability, or earnings smoothness variables.} 


With the missing definition of anomaly and the ad-hoc perturbations, HXZ's ``replication failures'' are more precisely described as ``insignificant ad-hoc trading strategies.''   Given that only about 26\% of HXZ's strategies  were shown to clearly produce statistical significance in the original papers (\citet{chen2020publication}), it's not surprising that roughly half of these ad-hoc strategies are insignificant, even when using equal-weighting.

\section{A Framework for Future Research}\label{sec:conclusion}
Outside of cross-sectional predictability, evidence on publication bias in asset pricing is scarce. Perhaps the best evidence on publication bias comes from  \citet{Goyal2008A}'s re-examination of 17 equity premium predictors and their on-going update, which adds another 29 predictors (\citet{Goyal2021A}).  

We conclude by using \citet{Goyal2008A} and  \citet*{Goyal2021A} to illustrate how our four facts can be applied to other fields.  We focus on real-time forecasts rather than simple regressions (\citet{goyal2003predicting}; \citet{Campbell2008Predicting}) for ease of comparison with the real-time results in Sections \ref{sec:facts}-\ref{sec:mis}.\footnote{The equity premium prediction literature typically refers to real-time prediction as ``out-of-sample tests.''  We use ``real-time prediction'' to differentiate from \citet{Mclean2016Does}'s out-of-sample tests (Section \ref{sec:facts-oos})} These papers provide the following evidence:
\begin{enumerate}
    \item \textbf{How replicable is equity premium predictability?}   Table 2 of \citet{Goyal2021A} suggests that replicability is rather poor.  Of 21 predictors that showed point estimates displaying real-time predictability in the original papers, only 10 have the same sign in \citet{Goyal2021A}'s replications.  Some of these differences are due to differences in the start dates of the real-time tests.  These results suggest that either author error or sampling error is has a meaningful impact in Equation \eqref{eq:intro3}. Regardless of the exact driver, the reported results are a biased estimate of what a follow-up replication would find.
    
    \item \textbf{Does equity premium predictability persist out-of-sample?} Predictability consistently weakens out-of-sample.  But to isolate publication bias effects, one would like to see predictability measures that zoom in on time periods near the original samples, a la \citet{Mclean2016Does}. Clear evidence on this question is currently lacking, and might come with large standard errors given the small number of observations relative to the cross-sectional asset pricing literature.
    
    \item \textbf{Are equity premium prediction p-values much smaller than 0.05?}   Tables 3-5 of \citet{Goyal2021A} strongly suggest the answer is no.  Of the 21 predictors that showed real-time predictability in the original papers, none is significant at the 1\% level in the replications.  Only six out of 21 are significant at the 5\% level.  While these tests use extended samples including post-sample data, these p-values are so large that it is unlikely that out-of-sample decay matters much.  Thus it seems  real-time prediction is readily explained by lucky draws, and sampling error in Equation \eqref{eq:intro3} is  large.
    
    \item \textbf{Are equity premium predictors weakly correlated?}  We downloaded data for \citet{Goyal2008A} from Amit Goyal's website, signed predictors, and computed correlations.  Correlations cluster around zero, and only a small fraction exceed 0.5. While this may sound surprising considering that all variables forecast the same left-hand side variable (the equity premium), it is consistent with the fact that combinations of forecasts predict the equity premium better than any single one of them (\cite{Rapach2010Out, Rapach2022AssetPricing}), and thus they must provide some uncorrelated variation. Low correlations suggest the p-value statistics in item 3 are good estimates of the underlying probabilities.
    
\end{enumerate}

Taken together, these results are consistent with publication bias being a major problem in real-time equity premium predictability. But one would like to see more rigorous answers to these questions, such as shrinkage estimates using the exact p-values.  One would also like to see the final version of \citet{Goyal2021A}'s on-going work, as well as follow-up studies by independent teams.   Nevertheless, these four facts show that publication bias certainly varies by the sub-field, and that other fields are not so unaffected by publication bias as cross-sectional predictability.  Indeed, the equity premium literature seems to acknowledge these issues, and has more recently focused on  methods that combine information from many individual predictors (\cite{Rapach2022AssetPricing}).


\pagebreak


\bibliographystyle{chicago}
\bibliography{literature.bib}

\clearpage


\end{document}